\documentclass[manuscript, screen]{acmart}
\AtBeginDocument{%
  \providecommand\BibTeX{{%
    \normalfont B\kern-0.5em{\scshape i\kern-0.25em b}\kern-0.8em\TeX}}}

\copyrightyear{2024} 
\acmYear{2024} 
\setcopyright{acmlicensed}\acmConference[LAK '24]{The 14th Learning Analytics and Knowledge Conference}{March 18--22, 2024}{Kyoto, Japan}
\acmBooktitle{The 14th Learning Analytics and Knowledge Conference (LAK '24), March 18--22, 2024, Kyoto, Japan}
\acmPrice{15.00}
\acmDOI{10.1145/3636555.3636918}
\acmISBN{979-8-4007-1618-8/24/03}
\usepackage{graphicx}
\usepackage{mathtools}
\usepackage{booktabs}
\usepackage{multirow}

\usepackage{enumitem}
\usepackage{subcaption}

\begin{document}

\title[Heterogenous Network Analytics of Small Group Teamwork]{Heterogenous Network Analytics of Small Group Teamwork: Using Multimodal Data to Uncover Individual Behavioral Engagement Strategies}

\author{Shihui Feng}
\email{shihuife@hku.hk}
\orcid{0000-0002-5572-276X}
\authornotemark[1]
\email{shihuife@hku.hk}
\affiliation{%
  \institution{University of Hong Kong}
  \city{Hong Kong}
  \country{China}
}

\author{Lixiang Yan}
\email{jimmie.yan@monash.edu}
\orcid{0000-0003-3818-045X}
\affiliation{%
  \institution{Monash University}
  \city{Clayton}
  \country{Australia}
}

\author{Linxuan Zhao}
\email{}
\orcid{}
\email{}
\affiliation{%
  \institution{Monash University}
  \city{Clayton}
  \country{Australia}
}

\author{Roberto Martinez Maldonado}
\email{roberto.martinezmaldonado@monash.edu}
\orcid{0000-0002-8375-1816}
\affiliation{%
  \institution{Monash University}
  \city{Clayton}
  \country{Australia}
}

\author{Dragan Ga\v{s}evi\'{c}}
\email{dragan.gasevic@monash.edu}
\orcid{0000-0001-9265-1908}
\affiliation{%
  \institution{Monash University}
  \city{Clayton}
  \country{Australia}
}

\begin{abstract}

Individual behavioral engagement is an important indicator of active learning in collaborative settings, encompassing multidimensional behaviors mediated through various interaction modes. Little existing work has explored the use of multimodal process data to understand individual behavioral engagement in face-to-face collaborative learning settings. In this study we bridge this gap, for the first time, introducing a heterogeneous tripartite network approach to analyze the interconnections among multimodal process data in collaborative learning. Students’ behavioral engagement strategies are analyzed based on their interaction patterns with various spatial locations and verbal communication types using a heterogeneous tripartite network. The multimodal collaborative learning process data were collected from 15 teams of four students. We conducted stochastic blockmodeling on a projection of the heterogeneous tripartite network to cluster students into groups that shared similar spatial and oral engagement patterns. We found two distinct clusters of students, whose characteristic behavioural engagement strategies were identified by extracting interaction patterns that were statistically significant relative to a multinomial null model. The two identified clusters also exhibited a statistically significant difference regarding students’ perceived collaboration satisfaction and teacher-assessed team performance level. This study advances collaboration analytics methodology and provides new insights into personalized support in collaborative learning. 

\end{abstract}





\maketitle

\section{Introduction}

Successful collaborative learning relies on the active participation of individuals \cite{dillenbourg1999you, kirschner2013toward}. Behavioral engagement can be considered the observed actions of individuals in collaborative learning processes \cite{fredricks2004school}. Understanding individual behavioral engagement in collaborative learning is critical for uncovering how learning occurs at the individual level, as well as for promoting cohesive and adaptive collaboration dynamics at the group level \cite{stahl2004building,kirschner2013toward}. The constitution of students' behavioral engagement often depends on the context and modality of the collaborative learning environments \cite{sharma2020multimodal}. Compared to large-scale online collaborative learning, such as class-level online forum discussions, small-group collaborative learning in physical settings provides a richer environment to engage students in the learning processes in various ways. In small-group co-located collaborative learning, individual behavioral engagement is multi-dimensional, manifesting in the multiple modes of communication and interactions during the learning processes, including oral communication, written exchange, and even spatial movements \cite{stahl2004building, yan2023socio}. During the learning process, individual students not only interact with their group members but also interact with the learning environment. These social and spatial interactions constitute the learning opportunities that determine the individual gains in collaborative learning \cite{dillenbourg1999you, kirschner2013toward,yan2023socio}. Uncovering individual behavioral engagement strategies in small-group collaborative learning is essential for providing personalized support, promoting self-regulation, and enhancing the effectiveness of collaborative learning \cite{hadwin2017self, strijbos2010developing}.

Previous studies on individual engagement in collaborative learning have been primarily conducted in wiki-based large-scale collaborative learning contexts \cite{hadjerrouit2016assessing, trentin2009using}. Among the studies focusing on individual engagement in small-group collaborative learning, log data were the most frequently used process data for analyzing individual engagement \cite{cen2016quantitative}. However, log data, such as the frequency and time spent on keystrokes, only record students’ interactions with the digital platforms used for assisting collaborative learning. In co-located small-group collaborative learning contexts, student behavioral engagement goes beyond what can be revealed solely from the log data \cite{gavsevic2015let}. Data that can capture students’ interactions with their group members and relevant learning objectives in the learning environments can enable a more comprehensive understanding of individual behavioral engagement in the learning processes \cite{riquelme2020you}. Therefore, multimodal process data, such as audio data of oral discussions and sensor data of spatial movements \cite{zhao2023mets}, are essential for gaining critical insights into individual behavioral engagement in physical collaborative learning settings. Little work has utilized multimodal data to analyze individual behavioral engagement in small-group collaborative learning. In this study, we aim to bridge this gap and propose a new heterogeneous tripartite network approach that can effectively analyze the interconnections among multimodal process data to uncover the characteristics of individual behavioral engagement strategies in small-group collaborative learning.  

For the first time, we introduce a heterogeneous tripartite network (HTN) approach to modeling individual behavioral engagement based on the interconnections among students, oral communication behaviors, and physical movements in co-located small-group collaborative learning. As opposed to traditional networks, which normally consist of only one type of node and edge, a heterogeneous network consists of multiple types of nodes and multiple types of edges \cite{shi2016survey}. For example, a bipartite network, which consists of two types of nodes and edges that only run between nodes from the two different types \cite{newman2018networks}, can be considered a special case of a heterogeneous network with only two distinct node sets. In a heterogeneous tripartite network (HTN), there are three types of nodes and three distinct sets of edges representing the connections between nodes from the three different possible pairings of distinct node types. The details of constructing the HTN in this study are provided in Section \ref{sec:htnconstruction}. Heterogeneous tripartite networks offer a unique and effective approach for analyzing the complex interconnections among students' social interactions and spatial movements in the collaborative learning process, which further enables us to identify the groups of students who share similar behavioral engagement patterns and uncover the specific strategies used by students within these groups. 

In this study, we collected the multimodal process data --- the oral communication data and spatial movement data --- for 15 teams, each consisting of four students participating in a small-group embodied team learning experience. Embodied team learning incorporates physical movement and interactive experience to facilitate students' learning in a group setting \cite{zhao2023mets}. The embodied team learning setting in this study was a set of team-based simulation activities within emergency healthcare. Students needed to move around in the learning space to complete the designed tasks collaboratively. Students' behavioral engagement in embodied collaborative learning was mediated mainly through their physical movement and oral communication. Therefore, the operational definition of students' individual behavioral engagement strategies in this study was students' interaction patterns with the spatial areas and oral communication types during the collaborative learning process. By modeling students' behavioral engagement with spatial positions and oral communication types using a heterogeneous tripartite network, we were able to identify (a) the individual students who shared similar behavioral engagement patterns during the learning process and (b) the association patterns among spatial movements and oral communications to reveal individuals' behavioral engagement strategies. As such, this study contributes a new methodological framework to analyze the interactions among multi-modal process data in collaborative learning, as well as provides new theoretical insights into understanding individual behavioral engagement in small-group co-located collaborative learning. The introduced heterogeneous tripartite network approach holds great promise in enriching the toolkit of multimodal learning analytics (MMLA) to analyze the underlying interconnections involved in collaborative learning processes. 

\section{Related work}

\subsection{Individual behavioral engagement in small-group collaborative learning}
Social constructivism and situated cognition provide the theoretical foundations for a comprehensive understanding of the individual dynamics during collaborative learning. Drawing from Vygotsky's foundational work, social constructivism elucidates the pivotal role that verbal interactions play in the construction of knowledge. It posits that through dialogue, learners actively engage in a mutual process of understanding and knowledge generation \cite{vygotsky1978mind}. From another theoretical perspective, the theory of situated cognition emphasizes the inherent connection between learning and its context. It argues that genuine understanding emerges from learners' active participation in authentic activities and that knowledge is both context-bound and application-oriented \cite{brown1989situated,lave1991situated}. Merging these perspectives suggests that a comprehensive understanding of collaborative learning requires a dual focus: one on the social interactions through which learners communicate and another on the environments in which these interactions unfold. Consequently, the current study endeavored to analyze both social and spatial behavioral engagement. The former offers insights into the cognitive and communicative nuances of learners, while the latter delves into the contextual aspects of their interactions. Such a dual analysis is instrumental in providing a nuanced, holistic grasp of the multifaceted dynamics at play in collaborative learning environments \cite{ferreira2021if}.

Students’ behavioral engagement is an indicator of active learning and is essential for effective knowledge co-construction in collaborative learning \cite{kirschner2013toward}. Previous studies on behavioral engagement in collaborative learning have mainly focused on the context of large-scale collaborative learning through online discussion forums \cite{hadjerrouit2016assessing, trentin2009using}. However, individual engagement methods can differ much between large-group and small-group collaborative learning, as well as between online and face-to-face (which we will refer to as \textit{offline}) settings \cite{strijbos2010developing}. In online collaborative learning settings, students’ participation is solely conducted online and their behavioral engagement can be relatively well reflected by log data such as the number of posts and the number of views for each post. However, in offline collaborative learning settings, in which group members collaborate in a face-to-face mode in a physical setting, individual participation is not limited to the use of digital platforms in the process. Rather, individual participation is largely reflected in students' oral discussions and physical interactions. Process data in multiple modalities is therefore essential for understanding behavioral engagement in collaborative learning. 

Existing studies on individual behavioral engagement have mainly focused on using online log data \cite{cen2016quantitative}, and little work has used multimodal data to analyze individual engagement in collaborative learning. Nasir et al. \cite{nasir2021many} employed video, audio and log data to characterize the learning behavioral profiles of dyads. Yan et al. \cite{yan2023characterising} categorized students into two groups based on their perceived stress, collaboration, and task satisfaction in small-group embodied teamwork. The differences in procedural and social behavior features between the different groups were analyzed using ordered network analysis (ONA). In the study by Yan et al. \cite{yan2023characterising}, the connections among multimodal process data entities were pre-defined by the behavioral codes, and network analysis was used to analyze the co-occurrence of the procedural and social behavior of individuals in the learning process, rather than to model the interconnections among multimodal behavioral data. Network analysis offers an effective way to analyze the structural characteristics of social connections among entities. However, analyzing the connections among multimodal process data in collaborative learning requires the analysis of multiple types of nodes and relationships. Therefore, in the current study, we introduced a new heterogenous tripartite network approach to model the interconnections among students, their communication behaviors, and spatial movements for analyzing individual behavioral engagement in small-group co-located collaborative learning. 

\subsection{Network analysis in collaboration analytics} 

A network is a collection of nodes joined by a set of edges, which can be used to model the interconnected components of real-world systems \cite{newman2018networks}. Network analysis offers a set of analytical tools for uncovering the hidden patterns of connections and structural characteristics within a given network \cite{borgatti2009network, newman2018networks}. Dado and Bodemer \cite{dado2017review} conducted a systematic literature review on the applications of network analysis in computer-supported collaborative learning (CSCL), finding that most studies utilized social network analysis to analyze the communication ties among students in CSCL. Dado and Bodemer also identified a lack of research analyzing the diverse relational ties involved in collaborative learning. In recent years, several studies have leveraged network analysis to analyze the relationships between students and learning artefacts in large-scale collaborative learning activities such as online forum discussions and collaborative annotations \cite{zhu2023understanding, poquet2023forum}. In addition, epistemic network analysis (ENA) offers an analytic tool to analyze the temporal co-occurrences of coded data in collaboration discourse \cite{shaffer2016tutorial}. ENA is proven to be an effective method for examining the connection structures between cognitive elements in user-generated content in collaborative learning processes \cite{gavsevic2019sens, lim2020students}. ENA has also been used with Social Network Analysis (SNA) to gain insights into both the social interactions and cognitive interactions among group members within CSCL \cite{gavsevic2019sens,swiecki2020isens}. 

Compared to large-size collaborative learning, the networks of social interactions in small-group collaboration tend to exhibit a simple topology, with three or four students forming a fully connected network. Given this simple topology, many global network centrality measures such as closeness or betweenness centrality are uninformative. Therefore, within the context of small-group collaborative learning, most of the studies applied network measures to analyze group-level performance \cite{zamecnik2023using}, rather than analyze individual engagement in small-group collaborative learning. However, network analysis can still be a powerful tool in analyzing these small-group collaborative learning settings, by allowing us to extract meaningful patterns in interactions beyond social interactions.

A heterogeneous network can be considered a generalization of a regular network in the sense that there are multiple types of nodes connected by multiple types of edges. In a regular one-mode network as commonly used in SNA for learning analytics \cite{gavsevic2019sens}, there is only one type of node connected by one type of edge. In a heterogeneous network, there can be multiple types of nodes and edges connecting nodes from different types. For example, bipartite networks, which consist of two distinct node sets, are a type of heterogeneous network \cite{newman2018networks, chen2023exploring}. There are a few studies deriving individual similarity networks by transforming a bipartite network capturing the associations of individuals and their generated content in online forums or collaborative online annotations \cite{zhu2023understanding,poquet2023forum}. For instance, Zhu and Chen \cite{zhu2023understanding} utilized two bipartite networks to model the associations between students and the keywords generated from their posts, and the associations between students and annotated sentences respectively. However, the specific bipartite network analysis techniques applicable to this large-group setting often do not provide meaningful results in the small-group setting. For instance, in a class-level collaborative annotation activity, the large number of students in the class could result in a wide array of different interests and interactions with the various parts of an article that the students are asked to read and annotate. This in turn leads to student-student networks that have more complex structures (e.g., bridging ties and long path lengths), and as such allows for the use of traditional SNA measures such as betweenness centrality, closeness centrality, or clustering coefficients, which can be used to understand the roles students take in discussion forums or the association of network measures as an operationalization of social capital with academic performance \cite{gavsevic2019sens}. 

In a small-group collaborative learning activity, if we use bipartite network analysis to model student-artifact interactions and then convert the bipartite networks into student-student networks, there is a high chance that the converted graph would be a fully connected network since there are only three or four student nodes. In addition, the bipartite network approach is effective in modeling the interactions of students with only one type of artefact in the learning process. However, in small-group co-located collaborative learning, understanding students' behavioral engagement requires the analysis of multimodal process data, including both social and spatial movement data. Therefore, in this study, we introduce a new heterogenous tripartite network that can model the relationships among three distinct types of nodes to fully leverage multimodal learning process data and capture more nuanced patterns in student engagement during collaborative learning processes.

\subsection{Research gap and research questions}
We summarize two research gaps aimed to be addressed in this study. Firstly, there is a lack of studies utilizing multimodal data to analyze individual behavioral engagement in collaborative learning. Previous studies primarily focus on analyzing team-level attributes and comparing the performance across teams in small-group collaborative learning \cite{zhao2023mets,yan2023sena}, but analyzing individual engagement in small-group collaborative learning using multimodal process data has not been a focus. 

Secondly, as the prior studies mainly used groups as the analysis unit in collaboration analytics research \cite{martinez2021you}, limited work has explored methods for analyzing individual engagement in collaborative learning using multimodal data. In this study, we propose a new heterogeneous network approach to model the interactions among students, communication behavioral, and spatial positions in small-group co-located collaborative learning. 

The proposed heterogenous tripartite network (HTN) models students' interactions with various communication behavior types and spatial locations in the embodied team learning process. Individual students could exhibit various engagement strategies in collaborative learning processes. It is critical to identify students who share similar behavioral engagement patterns with respect to oral communication and physical movement in the learning processes, which can pinpoint the distinct strategies employed by students and further support teachers in tailoring instruction to meet individual needs effectively. By identifying students who exhibit similar oral and spatial behaviors, we can uncover the common behavioral engagement strategies employed by students in a learning setting. This can provide in-depth insights into how students interact with their environment and peers, leading to the development of targeted interventions, improved teaching practices, and enhanced student outcomes. In summary, the two research questions leading this study are: 
\begin{itemize}
    \item[] \hspace{-1em}\textbf{RQ1:} To what extent can we group students who share similar behavioral engagement strategies with respect to communication behavior and spatial movements using a heterogeneous tripartite network approach? 
    \item[] \hspace{-1em}\textbf{RQ2:} What are the common spatial and communication engagement strategies among students in each group identified in RQ1? 
\end{itemize}

\section{Method}
\subsection{The learning setting}
We employed an existing dataset collected from a simulation-based learning scenario that was designed to support the development of students' teamwork skills in an emergency situation \cite{zhao2023mets}. This learning scenario, which was held over four weeks in 2021, was a compulsory course component of the Bachelor's of Nursing program at a university in Australia. In the simulation, each of group of four students completed various tasks assigned to four manikin patients in a specialized classroom simulating a hospital ward. Teachers monitored the students behind a one-way mirror and indirectly controlled the progress by managing the manikin patients. No intervention from teachers was made during this simulation. 

The whole simulation can be summarised with four phases: 
\textbf{(Phase 1)} Two students (primary nurses) entered the simulation room and received the medical information of the four patients from a teacher acting as a doctor. 
\textbf{(Phase 2)} After receiving all the necessary information, the primary nurses collaborated to make a plan, completing the tasks assigned to the four manikin patients. A few moments later, one manikin patient showed the signs of body deterioration. The primary nurses should recognize such signs and help this patient and then call the medical emergency team to allocate the two other students (i.e., secondary nurses) and an emergency doctor (role-played by a teacher) to help. \textbf{(Phase 3)} The secondary nurses entered the room to help. The primary nurses should provide handover information to the secondary nurses to involve them to help on resolving the emergency. These four nurses should collaborate to help the deteriorating patient (primary tasks) while completing tasks on other patients (secondary tasks). \textbf{(Phase 4)} After completing specific tasks, an emergency doctor entered the room and diagnosed the deteriorating patient. The nurses should communicate efficiently with the doctor to provide the necessary medical information for the doctor's diagnosis. The simulation ended after addressing the cause of deterioration. The study focused on Phases 3 and 4, which are those in which the four students were closely collaborating with each other. 

\subsection{Apparatus and Data Collection}
\label{sec:appartus-data-collection}
The dataset includes audio recordings, spatial coordinates, video, and team assessment data from the learning setting.
We used Xiaokoa portable wireless headset microphones to capture the voices of each consenting student. These microphones were connected to a multi-channel audio interface (TASCAM US-16$\times$08) to synchronize the audio streams and save them into individual audio files on a computer.
Each consenting student was provided with a waist bag containing a positioning sensor (Pozyx creator toolkit) to collect their spatial coordinates.
In relation to video data, we deployed a Jabra 180-degree camera to record a video for each session in which all four students provided us with informed consent.
These three modalities of data (audio, spatial positions, and video) were automatically synchronized by a customized software system. Last, we collected teachers' evaluations of students' team performance with respect to their collaboration and communication, as well as administered a post-survey to collect data on students' satisfaction with their team collaboration. The post-collaboration student survey included a single-item question that asked participants to rate their satisfaction with their team collaboration experience on a scale of 1 to 7.  Teachers' evaluations of team performance were collected through a questionnaire. The questionnaire was designed to evaluate three learning objectives with respect to teams' collaboration and communication performance, including teamwork effectiveness, communication effectiveness, and role comprehension. We used a 7-point Likert scale to assign scores to each of these learning objectives. The 15 teams were divided into two groups based on their scores: seven teams with low performance and eight teams with high performance, using the median value as the dividing point. This division was made to help us characterize the strategies that were identified as the result of the analysis performed in response to RQ1 (see Section \ref{sec:dataanalysis} for details).   

Using this data collection configuration, we analyzed multimodal data from the 15 student teams, each consisting of four students. Since there were two teams that only involved three students in phases 3 and 4, the analysis in this study was conducted on 58 students.

\subsection{Spatial movement and communication behaviors processing}

\subsubsection{Spatial and audio data processing}
The coding for spatial movement and communication behaviors was done at an utterance (i.e., turn of talk) level. To capture students' movement behaviors, we utilized a Python voice activity detection library (webrtcVAD) to detect the start and end points of students' utterances, from which the corresponding audio clips were then extracted. These extracted clips were subsequently transcribed by a third-party manual transcription service.
Next, we divided the simulation room into nine critical spatial areas related to students' tasks, namely, beds 1-4, three IV cabinets (IV cabinet uncolored, IV cabinet blue, and IV cabinet red), a phone area, and other non-working areas (other areas). Bed 4 and the phone area were the primary working areas since the deteriorating patient was placed on bed 4 and the students needed to use the phone to call the emergency team, while the remaining areas except  `other areas' were the secondary working areas. We used spatial coordinates to pinpoint the spatial areas within which a student produced each utterance.
Finally, we organized the transcribed utterances into dialogues referring to the spatial areas where the utterances were produced. Typically, the utterances produced in the same spatial areas would be organized into the same dialogue. To ensure all utterances were organized to correct dialogues, we manually checked through all the data according to recorded videos.

\subsubsection{Coding embodied team communication behaviors}
To analyze students' team dialogues, we employed the same coding scheme for embodied team communication in the previous study \cite{zhao2023mets}. The coding scheme adapts previous healthcare coding schemes \cite{miller2009identifying, riley2008nature} and a general teamwork theory framework \cite{bigfive}. Four higher-order teamwork constructs and eleven communication behaviors are included in this coding scheme. The first construct, \textit{shared leadership}, captures the communication when the students performed the team leader's tasks, such as coordinating the team or providing necessary information for new team members. The students commonly volunteered to perform such tasks, as no student was formally appointed as a leader \cite{carson2007shared}.
The second construct, \textit{situation awareness}, captures the students' identification and reaction to the emergency state that happened on one of four manikin patients \cite{riley2008nature}. 
The third construct, \textit{shared mental model}, captures the communication to reach a shared comprehension of situations \cite{vygotsky1978mind}.
The last construct, \textit{closed-loop communication}, captures the students' double-checking, agreement, and disagreement to ensure accurate communication of information or instructions \cite{miller2009identifying}. Each utterance could have multiple communication behavior codes. For example, an utterance of a student could be: \textit{"Her oxygen level is 89. We may need to call the emergency team"}. In this utterance, a student expressed \textit{information sharing} in "Her oxygen level is 89" and then suggested an \textit{escalation} in "We may need to call the emergency team". Two researchers coded 20\% of the dialogue data independently. These two researchers reached a Cohen's kappa score higher than 0.7 for each communication behavior, which indicated an acceptable agreement \cite{mchugh2012interrater}. The remaining 80\% of the data was coded by one researcher. 
The detailed definitions and examples for each communication behavior are provided in Table \ref{tab:scheme}.

\begin{table}[h]
\centering
\caption{Healthcare team communication coding scheme, including inter-rater reliability and frequency of each communication behaviour.}
    \label{tab:scheme}
\resizebox{0.85 \textwidth}{!}{%
\begin{tabular}{llllll}
\hline
\textbf{\begin{tabular}[c]{@{}l@{}}Teamwork\\ constructs\end{tabular}}                & \textbf{\begin{tabular}[c]{@{}l@{}}Communication \\ behaviours\end{tabular}}                                          & \textbf{Description}                                                                                                                    & \textbf{Example}                                                                                                                                                                                  & \textbf{Frequency} & \textbf{Kappa} \\ \hline
\multirow{3}{*}{\begin{tabular}[c]{@{}l@{}}Shared\\ leadership\end{tabular}}          & Task allocation \cite{riley2008nature}                                                               & \begin{tabular}[c]{@{}l@{}}Explicitly allocating\\ tasks to others or \\ self-allocating tasks.\end{tabular}                            & \begin{tabular}[c]{@{}l@{}}"Can you check \\ his heart rate?"\end{tabular}                                                                                                                        & 336 (12.7\%)       & 0.744          \\ \cline{2-6} 
                                                                                      & Planning \cite{stout1999planning}                                                                    & \begin{tabular}[c]{@{}l@{}}Showing the tasks \\ to be done for\\ provoking task \\ allocation.\end{tabular}                             & \begin{tabular}[c]{@{}l@{}}"We need complete\\ the discharge \\ form and do the \\ check list."\end{tabular}                                                                                      & 65 (2.5\%)         & 0.896          \\ \cline{2-6} 
                                                                                      & \begin{tabular}[c]{@{}l@{}}Provision of handover \\ information \cite{jorm2009clinical}\end{tabular} & \begin{tabular}[c]{@{}l@{}}Updating new-\\ coming participants\\ with necessary \\ information.\end{tabular}                            & \begin{tabular}[c]{@{}l@{}}"This patient \\ just finished her \\ hip-replacement, \\ she has..."\end{tabular}                                                                                     & 80 (3.0\%)         & 0.781          \\ \hline
\multirow{2}{*}{\begin{tabular}[c]{@{}l@{}}Situation\\ awareness\end{tabular}}        & Escalation \cite{brady2014qualitative}                                                               & \begin{tabular}[c]{@{}l@{}}Informing others the \\ situation goes beyond \\ the capabilities and \\ requesting extra help.\end{tabular} & \begin{tabular}[c]{@{}l@{}}"We need to call\\ the emergency \\ team and get a \\ doctor to help."\end{tabular}                                                                                    & 60 (2.3\%)         & 0.853          \\ \cline{2-6} 
                                                                                      & \begin{tabular}[c]{@{}l@{}}Situation \\ assessment \cite{holsopple2010enhancing}\end{tabular}        & \begin{tabular}[c]{@{}l@{}}Informing the change \\ in patients' status.\end{tabular}                                                    & \begin{tabular}[c]{@{}l@{}}"Her respiratory\\ rate is declining"\end{tabular}                                                                                                                     & 40 (1.5\%)         & 0.747          \\ \hline
\begin{tabular}[c]{@{}l@{}}Shared\\ mental\\ model\end{tabular}                       & \begin{tabular}[c]{@{}l@{}}Information \\ sharing \cite{van2011team}\end{tabular}                    & \begin{tabular}[c]{@{}l@{}}Proactively updating \\ information that was \\ not asked by others.\end{tabular}                            & \begin{tabular}[c]{@{}l@{}}"Her heart rate is\\ stable now."\end{tabular}                                                                                                                         & 488 (18.5\%)       & 0.744          \\ \cline{2-6} 
                                                                                      & \begin{tabular}[c]{@{}l@{}}Information \\ requesting \cite{alonso2012building}\end{tabular}          & \begin{tabular}[c]{@{}l@{}}Asking others a \\ question to collect \\ information.\end{tabular}                                          & \multirow{2}{*}{\begin{tabular}[c]{@{}l@{}}A: "How much oxygen\\  should we give?" \\ \textbf{(Information requesting)}\\ B: "Maybe starting\\ from four liters." \\ \textbf{(Responding to request)}\end{tabular}} & 556 (21.1\%)       & 0.794          \\ \cline{2-3} \cline{5-6} 
                                                                                      & \begin{tabular}[c]{@{}l@{}}Responding \\ to request \cite{alonso2012building}\end{tabular}           & \begin{tabular}[c]{@{}l@{}}Providing information \\ corresponding to a \\ previously asked \\ question.\end{tabular}                    &                                                                                                                                                                                                   & 312 (11.8\%)       & 0.804          \\ \hline
\multirow{3}{*}{\begin{tabular}[c]{@{}l@{}}Closed-loop \\ communication\end{tabular}} & Agreement \cite{hargestam2013communication}                                                          & \begin{tabular}[c]{@{}l@{}}Expressing agreement \\ on received information\\ or instructions.\end{tabular}                              & \begin{tabular}[c]{@{}l@{}}"Yes", "I agree", \\ "Okay."\end{tabular}                                                                                                                              & 636 (24.1\%)       & 0.858          \\ \cline{2-6} 
                                                                                      & Disagreement \cite{hargestam2013communication}                                                       & \begin{tabular}[c]{@{}l@{}}Expressing disagreement \\ on received information\\ or instructions.\end{tabular}                           & "No.", "I don't think so."                                                                                                                                                                        & 29 (1.1\%)         & 0.856          \\ \cline{2-6} 
                                                                                      & Checking-back \cite{hargestam2013communication}                                                      & \begin{tabular}[c]{@{}l@{}}Double-checking the\\ received information \\ or instructions.\end{tabular}                                  & \begin{tabular}[c]{@{}l@{}}A: "We need one \\ gram Ceftriaxone."\\ B: "one gram  Ceftriaxone?"\\ \textbf{(Check-back)}\end{tabular}                                                                      & 39 (1.5\%)         & 0.922          \\ \hline
\end{tabular}
}
\end{table}


\subsection {Heterogeneous tripartite network construction}
\label{sec:htnconstruction}
\begin{figure}[h]
  \centering
  \includegraphics[height=0.3\textwidth]{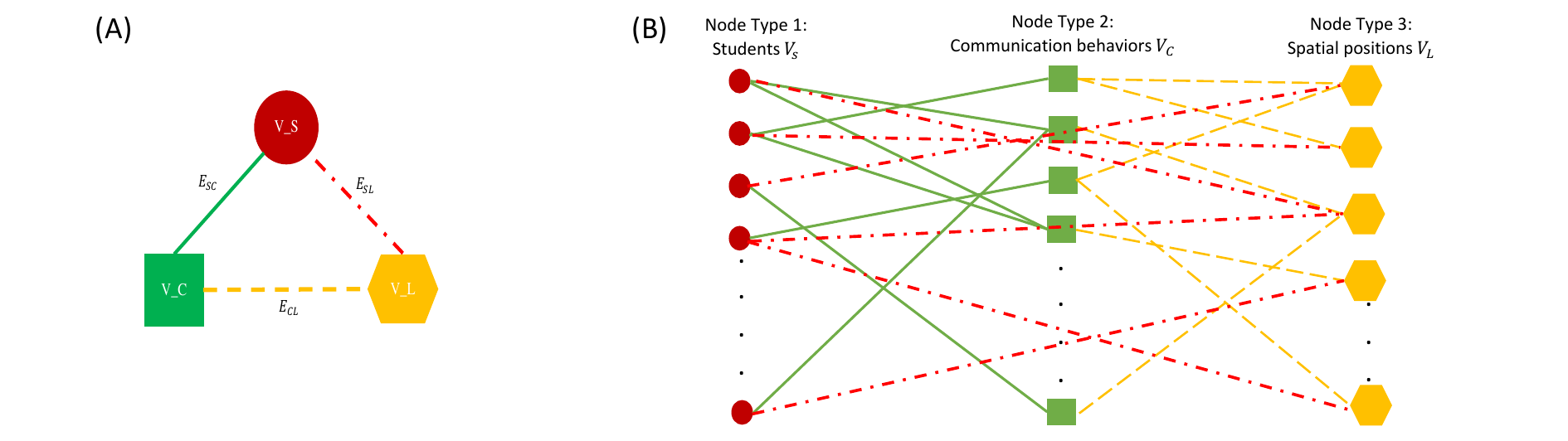}
  \caption{(A) A heterogeneous tripartite network $G=(\{V_S,V_C,V_L\},\{E_{SC},E_{SL},E_{CL}\})$ representing the individual engagement in spatial positions and communication behaviors during the embodied team learning process. Nodes $V_S$ represent students, nodes $V_C$ represent the coded communication behaviors, and nodes $V_L$ represent the spatial positions within the partition of the learning environment. $E_{XY}$ is the set of weighted, undirected edges representing the co-occurrence frequency among pairs of nodes in $V_X$ and $V_Y$. (B) An illustration of a heterogeneous tripartite network prior to any transformation  
  }
  \label{fig:fig1}
\end{figure}

In this study, we denote the heterogeneous tripartite network as $G = (V,E) = (\{V_S,V_C,V_L\},\{E_{SC},E_{SL},E_{CL}\})$, where $V_S$ is the set of $58$ students, $V_C$ is the set of 11 coded communication behaviors, $V_L$ is the set of nine areas partitioned in the learning space. Meanwhile, a weighted, undirected edge $(s,c)\in E_{SC}$ represents the number of times student $s$ engaged in communication type $c$ during the learning process; $(s,l)\in E_{SL}$ represents the frequency at which student $s$ engaged with the location $l$ within the learning environment; $(c,l)\in E_{CL}$ represents the frequency at which communication type $c$ was used within the location $l$ within the learning environment. In the constructed tripartite network, an individual student $s$ located at spatial position $l$ while engaging in communication behavior $c$ can be denoted as a triad $(s,l,c)$ accordingly. The set of triads an individual participates in summarizes the individual's spatial and communication engagement during the learning process, and the heterogeneous tripartite network $G$ represents the collection of individual spatial and communication engagement patterns in the embodied team learning process. Fig.~\ref{fig:fig1} is a schematic showing the structure of the heterogeneous tripartite network $G$ used in this study.

\subsection {Data analysis}
\label{sec:dataanalysis}

To answer RQ1, we first projected the tripartite network $G$ into a bipartite network $G_{S,(L,C)}$, whose edges run between the node set $V_S$ and the node set $V_C\times V_L$ (representing all possible pairs of locations and communication behaviors). The weight of each edge $(s,(l,c))$ in $G_{S,(L,C)}$ is the total weight (co-occurrence frequency) of engagement among a student $s$, location $l$, and communication behavior $c$, as counted by the number of triads with $(s,l,c)$ in the heterogeneous network. The bipartite projection $G_{S,(L,C)}$ provides a convenient and comprehensive representation of individual spatial and oral engagement strategies during the learning processes. It also allows for the application of a wide variety of tools for bipartite network analysis, including community detection methods. Community detection methods in bipartite networks cluster nodes of the same type based on their shared disassortative interaction patterns with the other node set \cite{karrer2011stochastic, yen2020community}. The first research question RQ1 is therefore naturally studied using bipartite community detection on the projected network $G_{S,(L,C)}$, as this can identify the clusters of students that have common engagement patterns in spatial positions and communication behaviors during the embodied team learning process. 

In this study, we applied stochastic blockmodeling to infer the clusters of students that have similar engagement patterns during the learning process. The stochastic blockmodel (SBM) is a generative model used for inferring block structure in networks \cite{holland1983stochastic, karrer2011stochastic}. There are several variants of SBM for improving its performance on real-world networks \cite{yen2020community,karrer2011stochastic, tarres2019tensorial}. In order to accommodate bipartite network structure and identify the optimal number of clusters in a nonparametric fashion, we employed the algorithm by \cite{peixoto2014efficient} for the degree-corrected stochastic blockmodel, which accounts for heterogeneity in the connection propensities of the nodes in its generative model for community structure. This algorithm identifies the optimal number of clusters and node assignments as the ones that minimize the description length of the model. To allow for the possibility of a null result---i.e., that there is no clustered structure in the network---we allowed the algorithm to sample the trivial partition of a single community during the optimization process. The optimal number of clusters was automatically determined by the algorithm minimizing the description length of the bipartite SBM, rather than by a pre-defined hyper-parameter. The algorithm returned two clusters of students as the optimal partition of the network, which demonstrates that, from an information theoretic perspective, there was sufficient evidence of clustered structure in the bipartite network. 

In order to support the interpretation of the two identified clusters, we also conducted an exploratory analysis to compare the post-activity collaboration satisfaction of students, as well as their affiliation with high or low-performing teams, between the two identified clusters. We report the analysis methods for comparing the differences between the two distinct identified groups here. To analyze the differences in students' perceptions of satisfaction between the two clusters, we employed a Mann-Whitney U test. This non-parametric test is suitable for comparing two independent groups when the data is not normally distributed or when the assumptions of parametric tests are not met \cite{mcknight2010mann}. To analyze the differences in students' associations with high and low performing groups (as defined in Section \ref{sec:appartus-data-collection}) between the two identified clusters, we employed Fisher's exact test, which is suitable for examining the association between two categorical variables in a contingency table with a limited sample size \cite{kim2017statistical}.  

To answer RQ2, we constructed two bipartite projections, $G^{(1)}_{L,C}$ and $G^{(2)}_{L,C}$, where the projection $G^{(i)}_{L,C}$ represents the co-occurrence patterns between location nodes $l$ and communication behaviors $c$ when restricted to the students within the cluster $i$ obtained using the blockmodeling in RQ1. More specifically, if $w_{s,l,c}$ is the number of triads among student $s$, location $l$, and communication behavior $c$ within the heterogeneous tripartite network $G$, then the weight of an edge $(l,c)$ in the projected graph $G^{(i)}_{L,C}$ corresponding to cluster $i$ is given by
\begin{align}
w^{(i)}_{l,c} = \sum_{s\in i}w_{s,l,c}.    
\end{align}
To identify the significant location-communication behavior characteristics within each cluster, we identified edges $(l,c)$ in each projected network that had statistically significant edge weights relative to a randomized null model. In the null model, all edges in the bipartite network $G^{(i)}_{L,C}$ are placed at random while fixing the degree of each location node $l$, which creates a maximally random topology while preserving heterogeneity in the location frequencies. This allows for a more stringent test of significance compared to using a null model that does not fix degrees of either node set. Edges $(l,c)$ whose weights are significantly larger than expected in this null model represent location-communication pairs that occur more often than expected given the overall interaction frequencies throughout the network. In this model, the marginal distribution of any particular edge weight $(l,c)$ in $G^{(i)}_{L,C}$ is given by a Binomial distribution with $k^{(i)}_l$ trials and success probability $1/\vert V_C\vert$, where 
\begin{align}
k^{(i)}_l=\sum_{c\in V_C}w^{(i)}_{l,c}    
\end{align}
is the weighted degree of location node $l$, and $\vert V_C\vert=11$ is the number of communication behaviors coded in the study. The threshold for an edge to be considered statistically significant at the $\alpha=0.05$ significance level can then be computed using the inverse survival function of the Binomial distribution \cite{yanagimoto1989inverse}, and all edges with weights below the threshold can be removed to obtain a sparse set of significant interactions that characterize the location-communication patterns within cluster $i$.

\section{Results}
\subsection{Student clusters based on the similarity of behavioral engagement strategies (RQ1)}

\begin{figure}[h]
  \centering
  \includegraphics[width=\textwidth]{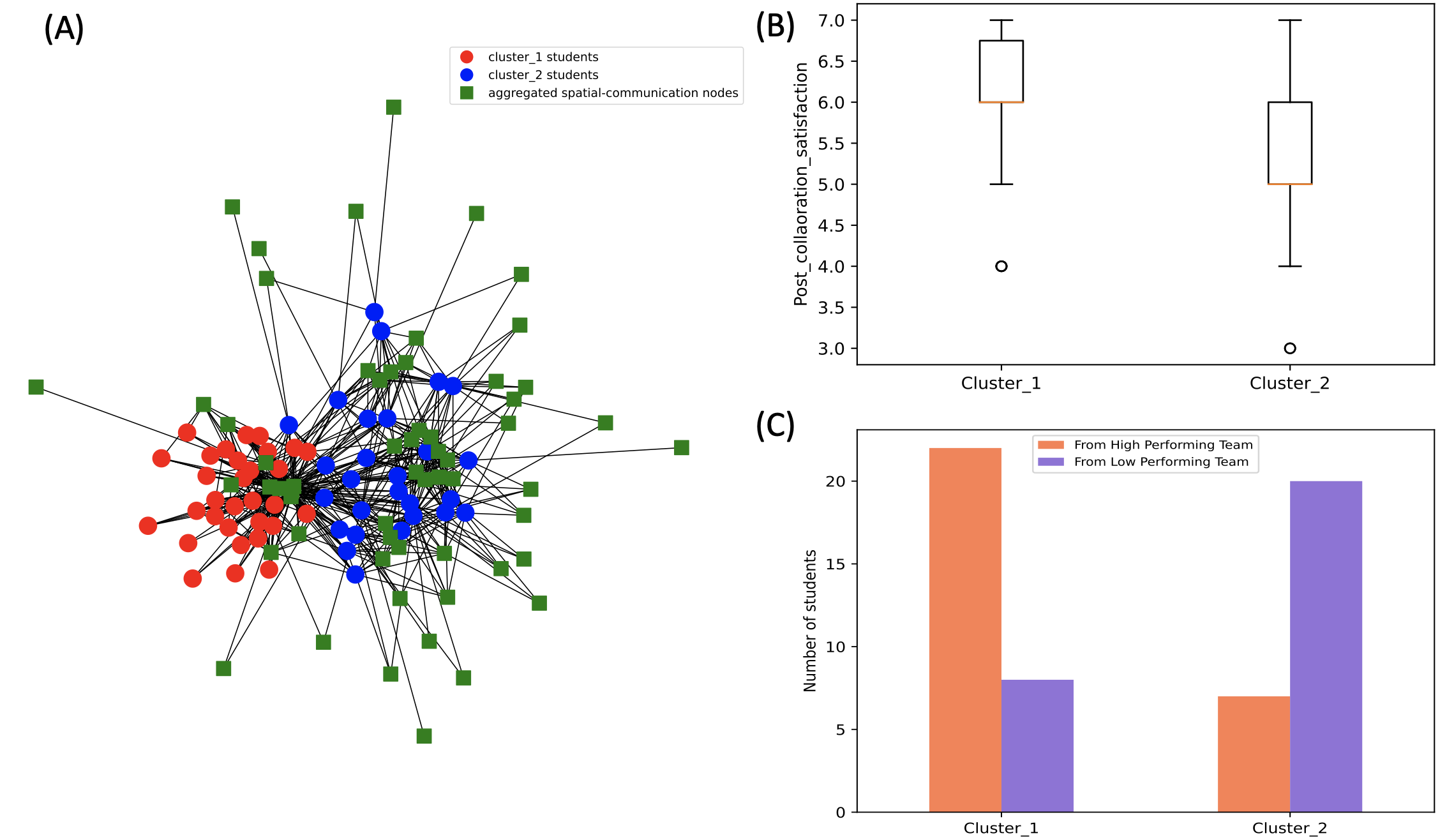}
  \caption{(A) Two clusters of students inferred with the bipartite stochastic blockmodel (SBM) \cite{yen2020community}. Circular nodes represent students in the studied cohort, while square green nodes represent the aggregated location-communication behavior pairs. Red and blue nodes indicate the two inferred clusters of students which exhibited distinct engagement patterns from each other. (B) Boxplots of the perceived post-collaboration satisfaction of the two clusters of students (medians in orange), which exhibit a statistically significant difference according to a Mann-Whitney U test. (C) Bar plot showing the number of students in each cluster that were a part of high and low performing teams. The clusters exhibit a statistically significant difference in their distributions of students' team performance according to a one-tailed Fisher's Exact test. }
  \label{fig:fig2}
\end{figure}

Two clusters of students with distinct behavioral engagement patterns were identified based on the stochastic blockmodeling analysis. Cluster 1 consists of 31 students and Cluster 2 consists of 27 students. Figure~\ref{fig:fig2}-A presents a visualization of the student clusters (red and blue nodes) and their connections with the nodes representing aggregated location-communication pairs (green nodes). There is a noticeable difference in how the green nodes (location-communication pairs) connect to each cluster, a pattern which is highlighted by the 2D embedding positions of the nodes in the spring layout used for the network visualization. This suggests that students in each cluster tend to engage with different combinations of oral communication types and spatial positions. The detailed differences in the behavioral engagement patterns captured by the location-communication pairs are reported in the next subsection.  

In order to gain further insights into the differences of students in the two clusters, we conducted an exploratory analysis here to analyze the difference in students' perceptions of the collaborative learning experience between the two groups based on their post-collaboration survey, as well as the difference in their affiliation with high or low performing teams based on the teachers' evaluation. Based on the Mann-Whitney U test result, we found a statistically significant difference in students' satisfaction with team collaboration between the two clusters ($U = 539$, $p = 0.025$, $r = 0.66$), with students in Cluster 1 ($\text{median}=6$) exhibiting higher post-survey satisfaction levels with large effect size than students in Cluster 2 ($\text{median}=5$). In Fig.~\ref{fig:fig2}-B, we plot the distribution of students' satisfaction levels for the two groups as boxplots to visualize these differences. We also found that there is a statistically significant association between cluster and team performance of students based on Fisher's exact test, with Cluster 1 containing significantly more students coming from high-performing teams (one-tailed, $p=0.0004$). The odds ratio (OR) for the association was estimated to be 7.86 (95\% CI: 2.41-25.61), indicating that the odds of students from a teacher-assessed high-performing team were 7.86 times higher among students in Cluster 1 compared to those in Cluster 2. In addition, as general rules of thumb, the odds ratio over 3.0 is considered a large effect size \cite{haddock1998using}. We plot the contingency table used for the Fisher's exact test as a bar plot in Fig.~\ref{fig:fig2}-C to visualize the difference in students' affiliated team performance in each cluster.

\subsection{Common behavioral engagement patterns for each student cluster (RQ2)}

Fig.~\ref{fig:fig3} below presents the common behavioral engagement patterns for each student cluster. We find that Cluster 1 students transitioned among five spatial positions during the learning process, including the two primary working areas (`phone' and `bed 4') and three secondary working areas (`bed 1', `bed 2', and `bed 3'). 
Although students in Cluster 1 engaged with all 11 communication behaviors, there were only six (location, communication behavior) pairs with which students in Cluster 1 engaged more than one would expect in the randomized null model. These include (`bed 4', `responding to request'), (`bed 4', `information requesting'), (`bed 4', `agreement'), (`bed 4', `information sharing'), (`bed 4', `task allocation'), and (`bed 2', `information requesting'). Students from Cluster 1 had a clear focus on the primary working area `bed 4' based on the analysis results. Various communication behaviors occurred in this location, which mainly focused on the category of shared mental model related to information sharing, requesting, and responding. In addition, task allocation, a type of communication code of shared leadership, was also conducted in the primary working area (bed 4). Bed 2, as the only secondary working area with which students in Cluster 1 had significant engagement, was associated with information requests. In summary, the behavioral engagement strategies of students in Cluster 1 exhibited a clear focus on the primary working areas and communication behavior related to information exchange in their learning processes. 

In contrast, students in Cluster 2 transitioned among a broader range of spatial positions, in addition to the primary working areas and the three secondary areas where students in Cluster 1 engaged (see Fig. 3). Students in Cluster 2 also engaged with the two IV cabinet areas and the other non-working area. There were 24 significant edges in Cluster 2, among which 20 connections were with the secondary working areas (`bed 1', `bed 2', `bed 3', `IV cabinet blue', and `IV cabinet yellow'), and four connections were with the primary working area (`bed 4'). The types of communication behavior that occurred mainly centered around the category of shared mental model, including information sharing, information requesting, and responding to requests, as well as agreement. In contrast to Cluster 1, task allocation occurred at both primary and secondary areas for the students in Cluster 2, and the students engaged actively with information sharing, requesting, and responding at the secondary working areas. In summary, the behavioral engagement strategies of the students in Cluster 2 exhibited more active physical movements across various primary and secondary working areas than in Cluster 1, and communication behavior related to information exchanges and task allocations occurred at a more diverse range of working areas than in Cluster 1, in which the students had a very clear focus on the primary working area.

It is intriguing to connect these two distinct identified engagement strategies with the variation in students' collaboration satisfaction and team performance across the two clusters. Students from Cluster 1, who had less variation in spatial movement and a clear focus on the primary working areas, had a higher level of post-collaboration satisfaction than the students from Cluster 2, who had more diverse spatial movements among the secondary working areas and links with a wider variety of communication behaviors. Meanwhile, most of the students from Cluster 1 were from high-performing teams according to teacher evaluations. This finding suggests a new theoretical hypothesis regarding the diversity of multimodal engagement patterns, subjective collaboration satisfaction, and performance level in collaborative learning. Based on the analysis results in this exploratory study, we observed that students with a higher spatial focus and consistent communication behavioral engagement patterns had a higher level of post-collaboration stratification and were more likely to be affiliated with a high performing team than students with a wider range of spatial movement and more diverse communication patterns.

\begin{figure}[h]
  \centering
  \includegraphics[width=\textwidth]{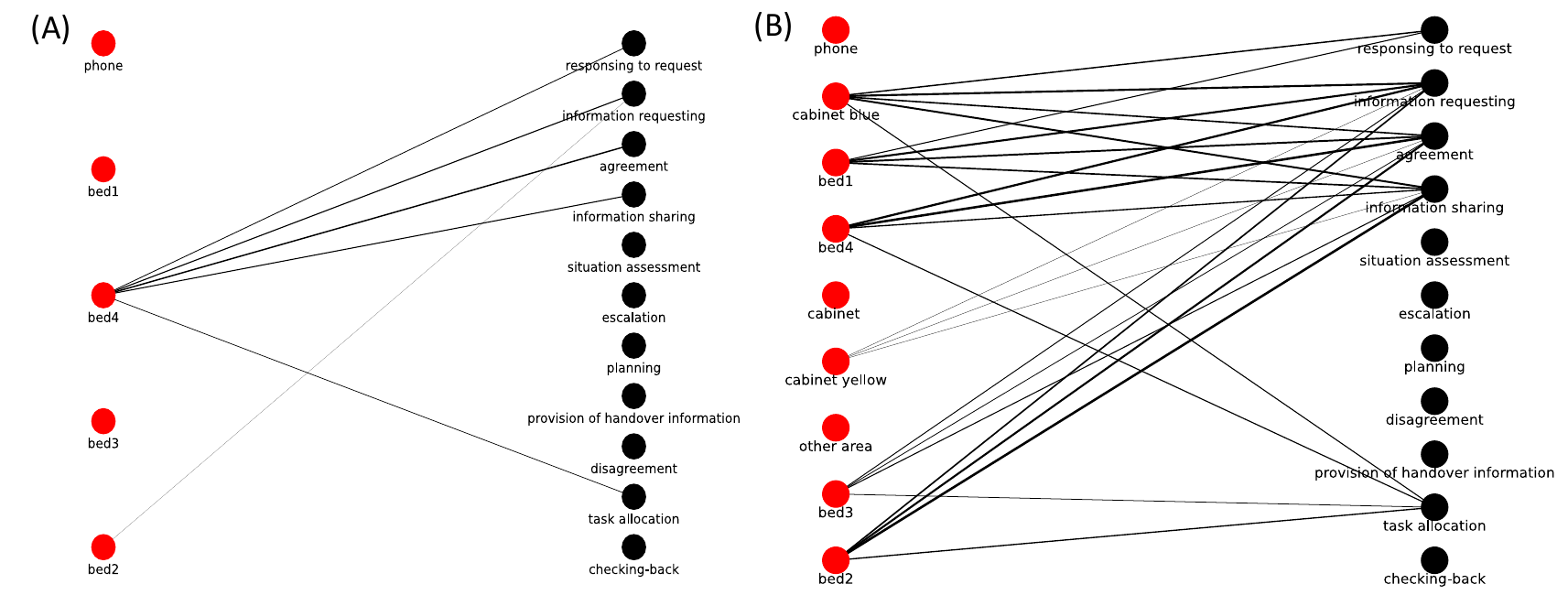}
  \caption{(A) Projected bipartite network of edges between locations and communication behaviors for students in Cluster 1, keeping only edges that were deemed statistically significant at the $\alpha=0.05$ level compared to a randomized null model (see Sec.~\ref{sec:dataanalysis}). (B) Projected network for Cluster 2 constructed using the same procedure. }
  \label{fig:fig3}
\end{figure}

\section{Discussion}

In this study, we analyzed individuals' behavioral engagement strategies with respect to spatial positions and oral communication types during small-group embodied team learning using a heterogeneous tripartite network approach. The results revealed that there were two clusters of students that shared similar individual engagement strategies with locations and communication types. The first cluster had a stronger focus on primary working areas with fewer spatial transitions during the collaboration, and mainly engaged in communication patterns related to information sharing, requesting, and responding at the primary working area. On the other hand, the students in the second cluster engaged in a greater number of spatial positions in secondary working areas in addition to the primary working area, and the associations between spatial positions and communication types were more diverse. In contrast with the students in the first cluster, students in the second cluster engaged in more communication behaviors at multiple secondary working areas such as information requesting, responding to requests, and information sharing. These findings between students' spatial and communication engagement patterns resonate with prior works that aim to bridge the connections among different theoretical perspectives of collaborative learning \cite{ferreira2021if}. Specifically, the current findings indicate the importance of analyzing collaborative learning through a holistic approach that combines social constructivism \cite{vygotsky1978mind} with situated cognition \cite{brown1989situated}. This combined approach can be particularly valuable in understanding students' individual engagement in physical learning spaces, where the types and patterns of their oral communication may depend on their spatial behaviors that are deeply connected to the learning design of the different areas within the learning space \cite{fernandez2021modelling}.

We also found that students in the two clusters had a significant difference in their perceived satisfaction regarding the collaborative learning experience, and that there was a significant difference in students' affiliation with high and low performing teams between the two clusters. Students with a strong focus on the primary working area and communication types connected to the primary working areas (Cluster 1) tended to have a higher level of satisfaction towards their collaborative learning experience, compared to the students with a higher level of spatial movement and more pairs of associations between spatial positions and communication behavior types (Cluster 2). This result leads to an interesting theoretical hypothesis for understanding individual engagement, which is that less physical movement and more stable behavioral engagement patterns in oral communication and spatial positions would lead to a higher level of satisfaction and better performance. We suggest future studies to further test this result experimentally and to analyze the underlying theoretical mechanism. Such findings also extend beyond prior works that focused on analyzing team-level engagement \cite{zhao2023mets} or individual-level spatial behaviors \cite{yan2023characterising, riquelme2020you}. Specifically, prior work that merely used spatial behaviors was unable to identify any statistical differences between students with low and high collaboration satisfaction \cite{yan2023characterising}, whereas such differences surfaced when using a multimodal approach when combining spatial behaviors with oral communication. This novel finding illustrated the advantage of using multimodal learning analytics over a single modality as different modalities may contain unique insights for understanding the differences in students' learning experiences, especially in complex learning settings \cite{cukurova2020promise}.

To uncover these findings with multimodal engagement data, in this study we introduced the application of heterogeneous tripartite networks to model the relationships among students, spatial positions, and oral communication in small-group embodied team learning. This study demonstrated that the heterogeneous network holds great promise in analyzing the interconnections of collaborative learning process data. Heterogeneous tripartite networks (HTNs) provide multiple advantages over existing methods such as Epistemic Network Analysis (ENA). First, HTNs allow for multiple node sets to be analyzed at a time — for example, the students, communication behaviors, and spatial locations analyzed in this study. Second, the heterogeneous network approach considers all node sets at once in an equal fashion, allowing for the application of a wide range of network analysis tools adapted for labeled networks, such as the Bipartite Stochastic Block Model we explore in this work \cite{tarres2019tensorial, peixoto2014efficient}. Third, the heterogeneous network analysis approach naturally accounts for correlations among multiple disjoint node sets representing multimodal learning process data \cite{shi2016survey}, while accommodating analyses of any subset of node types through marginalization over the connections with node sets that are not of direct relevance, or combination of node sets using the Cartesian product (as done in this study). Finally, HTNs allow for flexible visualization of the data of interest that does not only depend on the latent similarity among only one set of nodes (such as the ENA analysis conducted in the studies \cite{zhao2023mets}). The heterogeneous tripartite network provides a versatile and generalizable analytical approach that can be applied to analyze individual- and group-level structural characteristics of the interconnections among multimodal process data in collaborative learning contexts.

\section{Conclusion}

This exploratory study introduces a new heterogeneous tripartite network approach to analyze individual behavioral engagement using multimodal data in small-group collaborative learning. The proposed approach allowed us to identify meaningful groups of students who followed distinct individual engagement strategies based on the similarities of associations of multimodal process data, which could be used for supporting personalized recommendations in collaborative learning. The constructed heterogeneous tripartite network structure also allows us to examine the statistically significant edges among students and process outputs according to a suitable null model. The differences in students' perceived collaboration satisfaction between the identified clusters shed new light on the behavioral and affective relationships in collaborative learning. 

There are potential avenues for future work related to this study. Firstly, as heterogeneous networks are able to accommodate any number of distinct node types, one can use a heterogeneous network approach similar to the one we present here in other multimodal learning contexts, where behavioural patterns beyond oral communication and spatial movement may be of interest. Secondly, this study's learning context is based on healthcare. Future research is suggested to apply the heterogeneous tripartite network methodology of this study to other collaborative learning contexts to explore students' behavioral engagement patterns. In general, understanding individual behavioral engagement with multimodal data and the proposed analytical framework is critical for supporting teachers to provide data-informed and personalized support to enhance the effectiveness of collaborative learning.




\end{document}